\documentclass[a4paper]{article}%[a4paper,9pt,francais] 
\usepackage[utf8]{inputenc}
\usepackage{amsmath,amsfonts,amssymb,amstext,latexsym}
\usepackage{authblk}

\usepackage{graphicx}

\usepackage{url}

\usepackage[a4paper,left=2cm,right=2cm,top=1.5cm,bottom=1.5cm]{geometry}

\newcommand{\be}{\begin{equation}}
\newcommand{\ee}{\end{equation}}
\newcommand{\bea}{\begin{eqnarray} \nonumber }%
\newcommand{\eea}{\end{eqnarray}}
\newcommand{\bi}{\begin{itemize}}
\newcommand{\ei}{\end{itemize}}

\newcommand{\cRM}[1]{\MakeUppercase{\romannumeral #1}} % Capital
\newcommand{\dd}{\mathrm{d}}
\newcommand{\e}{\mathrm{e}}
\newcommand{\ii}{\mathrm{i}}

\newcommand{\hbe}{\hbar_\mathrm{exp}}
\newcommand{\Wconc}{W_\mathrm{conc}}
\newcommand{\vpconc}{\vec{p}_\mathrm{conc}}
\newcommand{\Econc}{E_\mathrm{conc}}

\title{Equivalent quantum equations in a system inspired by bouncing droplets experiments}
\author{Christian Borghesi \thanks{christian.borghesi@protonmail.com} \\
\normalsize{Équipe BioPhysStat, Université de Lorraine, 1 boulevard Arago, 57070 Metz, France}}
\date{}

\begin{document}

\maketitle

\abstract{
In this paper we study a classical and theoretical system which consists of an elastic medium carrying transverse waves and one point-like high elastic medium density, called concretion. We compute the equation of motion for the concretion as well as the wave equation of this system. Afterwards we always consider the case where the concretion is not the wave source any longer. Then the concretion obeys a general and covariant guidance formula, which leads in low-velocity approximation to an equivalent de Broglie-Bohm guidance formula. The concretion moves then as if exists an equivalent quantum potential. A strictly equivalent free Schrödinger equation is retrieved, as well as the quantum stationary states in a linear or spherical cavity. We compute the energy (and momentum) of the concretion, naturally defined from the energy (and momentum) density of the vibrating elastic medium. Provided one condition about the amplitude of oscillation is fulfilled, it strikingly appears that the energy and momentum of the concretion not only are written in the same form as in quantum mechanics, but also encapsulate equivalent relativistic formulas.
}

%\keywords{Classical system \and Wave-particle duality \and Guidance formula \and Equivalent quantum equations \and Quantum and relativistic energy}
%\pacs{46.40.-f ; 45.50.Dd ; 03.65.-w}

%%%%%%%%%%%%%%%%%%%%%%%%%%
%%%%%%%%%%%%%%%%%%%%%%%%%%
%%%%%%%%%%%%%%%%%%%%%%%%%%
\section*{Introduction}
Quantum-like phenomena were observed for the first time within the last decade in classical and macroscopic experiments, in which droplets bounce and `walk' on a vibrating liquid substrate (initiated in~\cite{yc_walking05} and see \cite{bush_review15} for a review). Experiments in which a droplet, called a walker, is guided by the wave that it has generated remind us of the {\it pilot wave} suggested by de Broglie~\cite{ldb_ondeguidee1927} (see \cite{yc_wavepartduality11,bush_review15,bush_phystoday16} for a discussion in this context). Nevertheless, it seems tricky to formalise mathematically the bouncing droplet problems in order to obtain equations close to the ones of corresponding quantum systems. To deal with more convenient systems, from the mathematical point of view, we have recently suggested a classical and macroscopic system~\cite{masselotte} -- also inspired by a sliding bead on a vibrating string experiment~\cite{yc_sao1999}. The system consisted of a bead oscillator free to slide on an elastic medium, whose transverse wave obeys a Klein-Gordon-like equation. This approach was encouraging, for instance we retrieved a strictly equivalent free Schrödinger equation. At this step there were still problems, the system for example restrained all the possible stationary solutions (obtained in an equivalent quantum system) to only one.

To overcome these limitations we just modify the previous system as follows: a very high elastic medium density -- called in this paper a {\it concretion} -- replaces the previous bead oscillator. This small change strengthens the wave-particle duality of the system and provides major improvements. It is moreover interesting to note that the classical toy system studied in this paper also exhibits indirect resemblances to the more quantum ones: \cite{holland1,holland2,holland3} where an inhomogeneity interacts with a quantum wave; and \cite{durt_1,durt_2} -- also inspired by the bouncing droplets experiments -- in which peaked quantum solitons are studied. These quantum models are specifically reminiscent of the {\it double solution} theory suggested by de Broglie~\cite{ldb_tentative,ldb_interpretation1987}.

The paper is organised as follows. In the first section, we establish the equation of motion for the concretion and the wave equation. We then focus on conditions for which the concretion is not the wave source any longer. We evaluate energy and momentum of the concretion, derived from the corresponding densities of the system. In the second section, we are looking for equivalent quantum equations of the system, more precisely the free Schrödinger equation and equations about energy and momentum. Then we study equivalent systems, without external potential in this study, which are commonly considered in quantum mechanics handbooks.

%%%%%%%%%%%%%%%%%%%%%%%%%%
%%%%%%%%%%%%%%%%%%%%%%%%%%
%%%%%%%%%%%%%%%%%%%%%%%%%%
\section{The theoretical system proposed and its dynamics}%\label{sth}
The system studied in this paper is no more than the one in~\cite{masselotte}, apart from a very little change: the bead oscillator of mass $m_0$ becomes, here, a very high density of the elastic medium itself. This high elastic medium density, of mass $m_0$, is called in this paper a concretion. As will be seen later, this small change provides very interesting  and major improvements.

The main goal of the devised system in~\cite{masselotte} was to deal with the main characteristics of the bouncing droplets experiments (see~\cite{bush_review15} for a review), while maintaining a convenient formalism to identify quantum analogies (and differences). These experiments with macroscopic and classical systems, in which droplets/walkers are piloted and guided by the wave that they have previously generated at each bounce, exhibit behaviours related to the wave-particle duality and some quantum-like phenomena. These experiments remind us the pilot-wave suggested by de Broglie. It is important to note that quantum-like behaviours are associated with a specific property of the system: (since the system is near the Faraday instability threshold) each transverse perturbation at the surface of the bath tends to generate a harmonic oscillation at the Faraday pulsation at the location of the perturbation. The more this tendency (also related to the {\it memory} of the system) appears, the more can occur quantum-like phenomena~\cite{yc_path-memory-pnas10,yc_SO-eingenstates14}. This property was mimicked in our suggested system in the following manner: any element of the elastic medium had the tendency to support transverse harmonic oscillations at pulsation $\Omega_m$, resulting from a quadratic/harmonic potential. Hence, the vibrating liquid bath in the bouncing droplets experiments has two properties: to carry propagative (transverse) capillary waves and {\it memory} effect. This vibrating liquid bath was simplified and has become in our theoretical system an elastic medium which supports transverse waves, obeying to a Klein-Gordon-like equation. Indeed this kind of medium has two tendencies: to carry propagative waves {\it à la} d'Alembert and to support standing wave, here at pulsation $\Omega_m$. Moreover, in order to deal with a simple mathematical formalisation, the suggested system was inspired by another experiment, in which a bead is free to slide on a vibrating string~\cite{yc_sao1999}. In the end, the macroscopic and classical system involved in bouncing droplets experiments has been transformed in~\cite{masselotte} into a bead oscillator free to slide on an elastic medium which supports transverse waves obeying to a Klein-Gordon-like equation.

%%%%%%%%%%%%%
%%%%%%%%%%%%%
%%%%%%%%%%%%%
\subsection{Framework and formalisation}
%Fig1. 
\begin{figure}
\begin{minipage}[c]{0.7\linewidth}
\centering
\includegraphics[width=1.0 \columnwidth,clip=true]{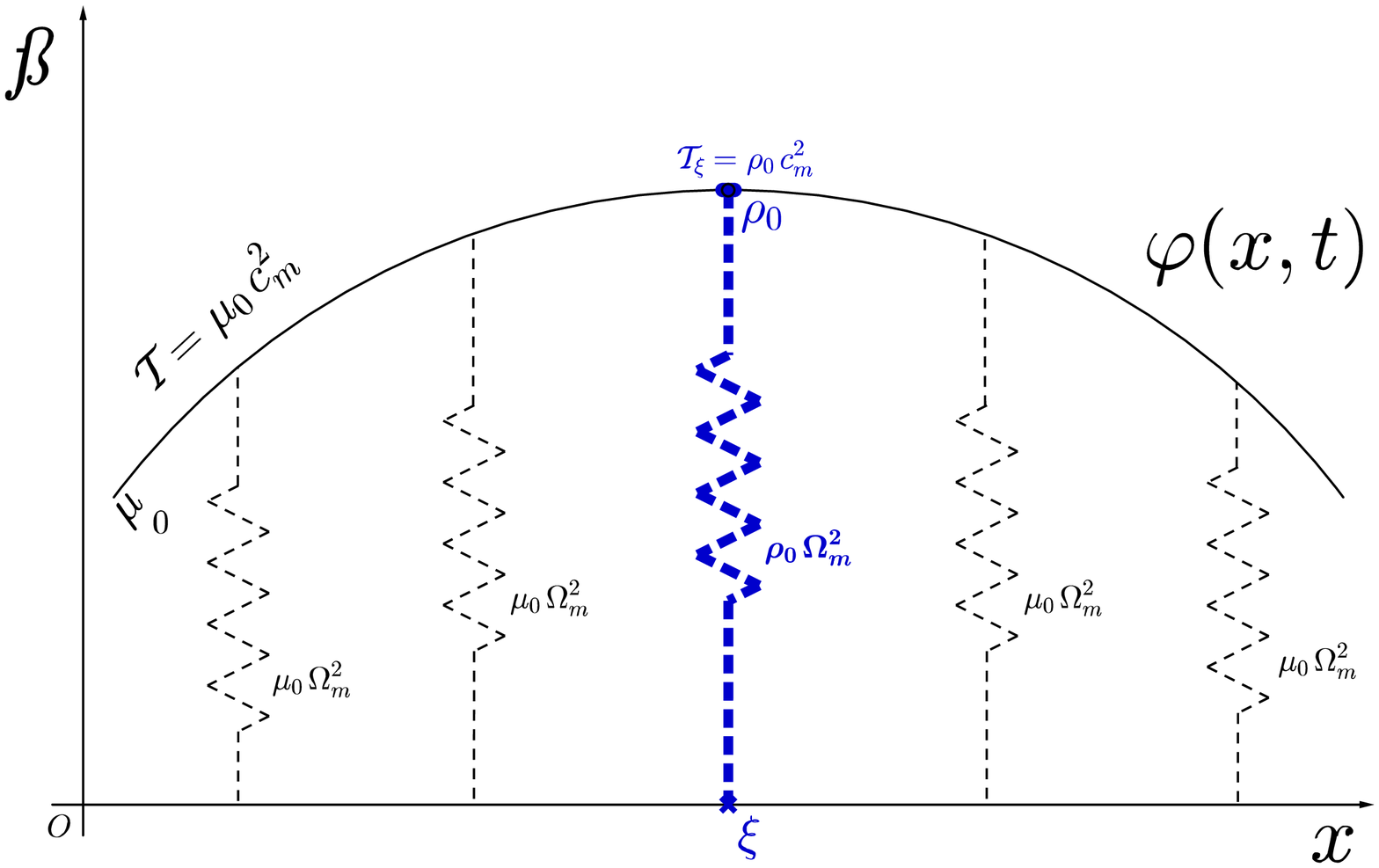}%\hfill %width=4cm, height=2.5cm,clip=true
\end{minipage}
\begin{minipage}[c]{0.28\linewidth}
\caption{\small Schematisation of the theoretical system, here in a 1D elastic medium and in the proper reference frame of the concretion. The harmonic potential of the medium (which mimic {\it memory} and Faraday pulsation in bouncing droplets experiments) is depicted by springs, whose stiffness per element of length are indicated. Features of the homogeneous elastic medium are represented in black, features of the concretion are in blue.}
\label{fig1}
\end{minipage}
\end{figure}

Even if main justifications and explanations about the suggested system are given in~\cite{masselotte}, let us again shortly describe this system -- which we would like to be implementable in practice. The system (see Fig.~\ref{fig1} for a schematic representation) is constituted as follows:
\bi 
\item[-] An homogeneous, isotropic, elastic medium of three dimensions (where $\vec{r}$ denotes its coordinates). (The spatial dimension is a free parameter for this system.) This elastic medium is subjected to a {\it tension} $\mathcal{T}$ and has a mass per element of volume $\mu_0$. The medium carries transverse waves. $\mathcal{T}=\mu_0\,c_m^2$, where $c_m$ is the propagation speed of the wave -- whose specific value is not important in this article. The transverse displacement at point $\vec{r}$ and time $t$ is $\varphi(\vec{r},t)$. (It is worth noting that the transverse displacement is a real number -- and in consequence $\varphi$ constitutes a scalar field -- as for example the interface height of the liquid bath in bouncing droplets experiments.)  The elastic medium tends to support transverse standing vibrations at pulsation $\Omega_m$. In the end, $\varphi(\vec{r},t)$ is thus a scalar field which obeys a Klein-Gordon-like equation -- when there is neither any source nor external excitation.
\item[-] The elastic medium contains an exceed mass heterogeneity, $m_0$, constant in time. It is important to notice that this exceed mass does not comes from an external mass, like a bead in~\cite{masselotte}, but comes from the elastic medium itself. We assume that the medium, which normally has a mass per element of volume $\mu_0$, carries a very concentrated heterogeneity of the medium itself: the concretion. We assume this heterogeneity as a point, at the location $\vec{\xi}$ at time $t$. The exceed mass density is written in the reference frame $\mathcal{R}_0$ proper to the concretion as:
\be \label{eth_rho0}
\rho_0(\vec{r}_0,t_0) = m_0\,\delta(\vec{r}_0-\vec{\xi}_0) \,,   
\ee 
where $\delta$ denotes the Dirac delta function. Lastly we assume that this exceed of elastic medium shares the same properties as the elastic medium itself. In other words the heterogeneity has the same properties as the homogeneous elastic medium times $\frac{\rho_0}{\mu_0}$. 
\item[-] The space is separated between the three-dimensional space of the elastic medium (which is called, for short, {\it perceivable space}) and an axis which is specific to the transverse vibrations. (In bouncing droplets experiments the horizontal surface bath denotes the perceivable space -- on which a droplet `walks' -- while the vertical axis is the transverse axis of oscillations; in the sliding bead on a vibrating string experiment~\cite{yc_sao1999} the perceivable space -- along which the bead slides -- is the axis of the string at rest while the axis of transverse oscillation is the vertical axis.) In this paper, this distinction of space is due to the fact that an observer looking at the dynamics of a point mass along a vibrating medium pays usually more attention to the sliding motion of the point mass than its transversal vibration -- as for example in bouncing droplets systems. Nevertheless, it could be of interest to mention that de Broglie has studied the possibility of an existing wave mechanics in presence of a fifth dimension hidden to our senses~\cite{ldb_5dim1927}. However we must keep in mind, that in our suggested classical system, the transverse oscillation velocity of the concretion and transverse displacements are measurable -- as the vertical velocity of a bouncing droplet and surface heights of the liquid bath. Anyway, recall that $\vec{r}$ specifies a point of the elastic medium at rest ({\it i.e.} the perceivable space) while $\varphi$ indicates a transverse displacement of the elastic medium. The perceivable velocity of the concretion, $\vec{v} = \frac{\dd \vec{\xi}(t)}{\dd t}$, the slope of the transverse wave, $\vec{\nabla}\varphi$, etc. are expressed in the perceivable space. (In the same manner, a walking velocity of a droplet is written in the 2D horizontal surface of the bath.) In the following, we do not use the term perceivable when referring to the location, the velocity, etc. of the concretion if it is not necessary.
\item[-] The elastic medium is thus not dissipative. Moreover, gravity, friction, non-linearities and other dissipative effects are neglected -- in theory...
\item[-] We only consider the interaction of the concretion with the elastic medium, as if nothing else was interacting with it. Since a Klein-Gordon-like equation governs the wave dynamics in the elastic medium and this equation is invariant by the Lorentz-Poincaré transformation (with $c_m$), the study is covariant with respect to the elastic medium (in particular the propagation speed of the wave $c_m$).
\item[-] Apart from the study of a linear cavity (cf. Section~\ref{sssim_lin}), calculations are made in a 3D elastic medium in the following. Indeed, the introduction of three spatial degrees of freedom for the (perceivable) space is necessary for developing a quantum analogy, but this spatial dimension is a free parameter, rather irrelevant, for this study. On the other hand, from an experiment point of view, a 1D or 2D (as for bouncing droplets experiments) elastic medium would suffice for the implementation of this theoretical system.
\ei

The system has only one Lagrangian density, the one of the wave. It contains a part due to the d'Alembert equation ($\propto [(\frac{1}{c_m}\frac{\partial\varphi}{\partial t})^2 - (\vec{\nabla}\varphi)^2]=\partial_\mu \varphi \, \partial^\mu \varphi$), and another part due to the tendency of the medium (coming from a quadratic potential) to support standing waves at pulsation $\Omega_m$. Including the concretion, the Lagrangian density of the system is thus written as
\be \label{eth_l}
\mathcal{L} = \frac{1}{2}\, \mathcal{T}\,\left(1\,+ \, \frac{\rho_0(\vec{r},t)}{\mu_0} \right) \left[\partial_\mu \varphi \, \partial^\mu \varphi  \,-\, \frac{\Omega_m^2}{c_m^2} \varphi^2 \right] \,. 
\ee 
(See Appendix~\ref{sa_comparison} for a basic comparison between this Lagrangian density and the one with the bead oscillator presented in~\cite{masselotte}.) As expected and discussed above, we notice that ($i$) this Lagrangian density is Lorentz-Poincaré invariant with respect to the considered elastic medium ({\it i.e.} the propagation speed of the wave, $c_m$) and ($ii$) the system is wave monistic (since the `particle' concretion is described by the transverse wave, $\varphi$, at the location of the high elastic medium density).

We also notice that the toy system suggested in this paper considers the concretion as a stable particle. In other words, the stability of the high elastic medium density (the concretion) is out of the scope of this paper -- as well as the stability of particles in non-relativistic quantum mechanics. Similarly, neither the question of bouncing droplets' stability nor very complex phenomena concerning interaction between a droplet and the liquid bath (see {\it e.g.} \cite{molacek1,molacek2}) are usually taken into account when walkers' dynamics is studied. Nevertheless let us evoke another very interesting approach (indirectly connected with our paper), also inspired by bouncing droplets experiments, in which dynamics of peaked quantum solitons -- due to a non-linear self-focusing potential of gravitational nature -- are studied~\cite{durt_1,durt_2}. According to this viewpoint, the concretion could be seen, for instance, as a simplification of more complicated phenomena. 

Finally, let us discuss the toy system presented here in the light of the de Broglie double solution program~\cite{ldb_tentative,ldb_interpretation1987} (see~\cite{fargue,double_sol_90ans} for current and well-thought overviews). Very briefly and in the scope of our study, the double solution program -- which is also a wave monistic theory -- distinguishes two waves: ($i$) the $u$ wave, which has a very high amplitude around a point, accounts for the particle and ($ii$) the $\psi$ wave, which is continuous, pilots the particle. In contrast to the double solution theory, the `particle' concretion is depicted by a very high elastic medium density rather than a very high amplitude of the wave. (This becomes possible in our theoretical system because the elastic medium is a material medium, which has a mass per element of volume.) The transverse wave, $\varphi$, associated to the elastic medium can thus be everywhere continuous. Furthermore, as seen below (cf. Section~\ref{ssth_guidance}), the transverse wave, $\varphi$, plays also the role of the ``pilot wave" -- devoted to $\psi$ in de Broglie-Bohm mechanics. We note nevertheless that our paper will not study another complicated and subtle behaviour, a possible very rapid motion of the concretion around its average location (cf. Section~\ref{ssth_guidance}). In the end, as this paper does not take into account neither the very precise behaviour of the concretion nor the stability of the concretion, this study could also be interpreted as a simplification (with the modification concerning the description of the particle) of the de Broglie double solution theory in a classical system.

%%%%%%%%%%%%%
%%%%%%%%%%%%%
%%%%%%%%%%%%%
\subsection{The equation of motion for the concretion and the wave equation}
The wave equation and the equation of motion for the concretion stem from the principle of least action (see Appendix~\ref{sa_equas} for more details). The equation of motion for the concretion is written as
\be \label{eth_mvt}
\frac{\dd}{\dd \tau}\left[\frac{m_0}{2}\,\left(\partial_\mu \varphi\partial^\mu\varphi - \frac{\Omega_m^2}{c_m^2}\,\varphi^2 \right) \, U_\alpha\right] = -\,m_0\,c_m^2 \;\partial_\alpha \varphi   \left(\square_m\varphi + \frac{\Omega_m^2}{c_m^2}\,\varphi \right) \,,
\ee
where $\square_m$ denotes the d'Alembert operator (specific to the elastic medium), $\tau$ the proper time of the concretion and $U_\alpha$ a covariant component of the 4-velocity of the concretion; and the wave equation as
\be \label{eth_ch}
\square_m\varphi \,+\, \frac{\Omega_m^2}{c_m^2}\,\varphi = -\,\frac{\rho_0}{\mu_0}\left(\square_m\varphi \,+\, \frac{\Omega_m^2}{c_m^2}\,\varphi \right) \, - \, \frac{1}{\mu_0}\left[\partial_\mu(\rho_0)\,\partial^\mu \varphi \right] \,.
\ee

The equation of motion shows that the concretion should be deflected by the transverse wave $\varphi$, in particular by its slope at the location of the concretion. Note that the term in brackets before the 4-velocity, $\frac{m_0}{2}(\partial_\mu \varphi\partial^\mu\varphi - \frac{\Omega_m^2}{c_m^2}\,\varphi^2)$, can be seen as a kind of effective mass which could oscillate in time. The wave equation can be interpreted as an inhomogeneous Klein-Gordon-like equation in presence of a wave source. The source is localised at the position of the concretion and depends moreover on its vibration.

%%%%%%%%%%%%%
%%%%%%%%%%%%%
%%%%%%%%%%%%%
\subsection{The concretion guidance formula \label{ssth_guidance}}
It is now interesting to investigate the special case in which the concretion is not the wave source any longer. In a mathematical point of view, this case is particularly significant since it allows us to establish an equivalent Schrödinger equation for the system (cf. Section~\ref{sssim_equa} and \cite{masselotte}). Inspired by bouncing droplets experiments, {\it e.g.}~\cite{yc_orbiting_source13,yc_SO-eingenstates14,yc_auto_orb16}, where stable orbits of walkers (which present some analogies with quantum states) result from self-organised phenomena between walkers and the wave at the surface of the bath, we assume that the concretion could not be the wave source any longer after a kind of self-adaptive phenomenon between the transverse wave $\varphi$ and the concretion. In this case, there is a strong relationship, or an intimate harmony, between the wave and the concretion. In particular, there is no longer back-reaction of the concretion on the wave $\varphi$. We have called this state {\it symbiosis} in~\cite{masselotte}. Hence, once this state is reached, an observer who focuses on the waves would be blind to the concretion. The concretion is in the wave and yet behaves as if it was not.

According to the wave equation~(\ref{eth_ch}), the concretion does not generate waves: ($i$) when the wave is governed by the Klein-Gordon-like equation and ($ii$) when 
\be \label{eth_ursymb}
\partial_\mu(\rho_0)\,\partial^\mu \varphi = 0 \,.
\ee 
(We call this equation, which has very interesting consequences in this toy system, the {\it symbiosis equation}.) According to the mass conservation, this equation means in the proper reference frame of the concretion that the slope of the transverse displacement, $\varphi$, is zero at the location of the concretion ({\it i.e.} $\vec{\nabla\varphi}(\vec{\xi},t)=0$ in $\mathcal{R}_0$). The concretion is thus located in its proper reference frame at a local extremum of the transverse displacement wave $\varphi$. Moreover, we see later (cf. Eq.~(\ref{eth_vGuid})) that this relation, in the low-velocity approximation, leads to an equivalent equation to the de Broglie-Bohm {\it guidance formula} -- which played a great role in his investigations~\cite{ldb_tentative,ldb_interpretation1987} and also in Bohmian mechanics~\cite{Q_th_motion,applied_bohmian}. 

In the reference frame $\mathcal{R}$ where the concretion has a velocity $\vec{v}$, the proper mass density is written as $\rho = \gamma_m\,\rho_0$, in which the Lorentz factor, $\gamma_m=1/\sqrt{1-v^2/c_m^2}$, is specific to the elastic medium. According to the mass continuity equation ({\it i.e.} here, $\frac{\partial \rho}{\partial t} + \vec{v} \cdot \vec{\nabla}\rho = 0$), Eq.~(\ref{eth_ursymb}) (where $\rho_0 = \frac{\rho}{\gamma_m}$) leads to:
\be \label{eth_ursymb2}
\frac{\rho}{c_m^2}\frac{\partial (\frac{1}{\gamma_m})}{\partial t}\frac{\partial \varphi}{\partial t}(\vec{\xi},t) - \frac{1}{\gamma_m}\left[\frac{\vec{v}}{c_m^2} \, \frac{\partial \varphi}{\partial t}(\vec{\xi},t) \: + \: \vec{\nabla}\varphi(\vec{\xi},t)\right]\cdot \vec{\nabla}\rho = 0 \,.
\ee
(This equation can be directly established using Eq.~(\ref{eth_ursymb}), $\rho_0(\vec{r},t)=\frac{m_0}{\gamma_m}\delta(\vec{r}-\vec{\xi}(t))$, $m_0$ is constant in time and $\frac{\partial \delta(\vec{r}-\vec{\xi}(t))}{\partial t}=-\vec{v}\cdot\vec{\nabla}\delta(\vec{r}-\vec{\xi}(t))$.) To make clearer the reasoning, we use a representation of a Dirac delta function as a limit of a derivable function.~\footnote{Let, for instance, the mass density expressed by a 3D Gaussian function, $\rho(\vec{r},t)= \frac{m_0}{(\sigma\sqrt{2\pi})^3}\exp[-\frac{(\vec{r}-\vec{\xi})^2}{2\,\sigma^2}]$, where $\sigma\rightarrow 0$. Hence, Eq.~(\ref{eth_ursymb2}) becomes $\left(\frac{\gamma_m}{c_m^2}\frac{\partial \frac{1}{\gamma_m}}{\partial t}\frac{\partial \varphi}{\partial t} + \left[\frac{\vec{v}}{c_m^2} \, \frac{\partial \varphi}{\partial t}(\vec{\xi},t) \: + \: \vec{\nabla}\varphi(\vec{\xi},t)\right]\cdot \frac{\vec{r}-\vec{\xi}}{\sigma^2}\right)\rho(\vec{r},t) = 0.$ This leads, when $\vec{r}=\vec{\xi}$, to $\frac{\partial}{\partial t} (\frac{1}{\gamma_m})=0$.} When $\vec{\nabla}\rho=\vec{0}$ ({\it i.e.} at the maximum of $\rho$), the above relation~(\ref{eth_ursymb2}) yields $\gamma_m$ constant in time. Hence, the relation~(\ref{eth_ursymb2}) implies the two following results: ($i$) $\gamma_m$ and then the speed $v$ of the concretion are constant in time (which is in agreement with the conservation of the energy a particle in the absence of external potential) and ($ii$)
\be \label{eth_symb}
\frac{\vec{v}}{c_m^2} \, \frac{\partial \varphi}{\partial t}(\vec{\xi},t) \: + \: \vec{\nabla}\varphi(\vec{\xi},t) \: = \: \vec{0} \,.
\ee %$\rho=m_0\delta(\vec{r}-\vec{\xi}(t))$
This relation, called in the following $\varphi$-guidance formula (recall $\varphi$ is a real number and denotes throughout this paper a transverse displacement and not a phase), plays a crucial role in this paper. It is worth noting that this relation is very general and is not restricted to the low-velocity approximation. Moreover Eq.~(\ref{eth_symb}) has a covariant form and (again) expresses that $\vec{\nabla\varphi}(\vec{\xi},t)=0$ in $\mathcal{R}_0$ ({\it i.e.} the gradient vector of the transverse wave $\varphi$ is zero at the location of the concretion in its proper reference frame).\\

Let us now write the $\varphi$-guidance formula~(\ref{eth_symb_mvt}) in the low-velocity approximation and in a more usual form ({\it i.e.} using a phase). It is convenient to introduce the modulating wave, $\psi$, which modulates the `natural' wave of the medium. As was described in~\cite{masselotte}, $\psi$ is specifically associated with the presence of the concretion in the elastic medium. Using the complex notation:
\be \label{eth_psi_def}
\varphi(\vec{r},t) = \mathrm{Re}\left[ \psi(\vec{r},t) ~ \e^{-\ii\,\Omega_m\,t}  \right] \,,
\ee
where $\mathrm{Re}[\cdots]$ denotes the real part. (According to this expression, $\psi$ naturally appears as a complex wave.) In the low-velocity approximation we get $|\frac{\partial \psi}{\partial t}| \ll \Omega_m |\psi|$ (cf.~\cite{masselotte}). In addition, it is convenient to write the modulating wave as
\be \label{eth_psi_FPhi}
\psi = F\:\e^{\ii  \Phi} \,,
\ee
where the magnitude $F$ and the phase $\Phi$ are two real functions. In the low-velocity approximation, the imaginary part from the $\varphi$-guidance formula~(\ref{eth_symb}) leads to
\be \label{eth_vGuid}
\vec{v} = \frac{c_m^2}{\Omega_m} \: \vec{\nabla}\Phi(\vec{\xi},t) \,.
\ee 
This equation is an equivalent of the {\it formule de guidage} established by de Broglie (Eq.~(26') in~\cite{ldb_ondeguidee1927}) and differently obtained later by Bohm~\cite{bohm1}. (See~\cite{yc_wavepartduality11} for an acute discussion between the two different viewpoints, the de Broglie {\it double solution} and the bohmian mechanics, in the context of walkers. In addition, it is interesting to notice that walkers do not follow an equivalent de Broglie-Bohm guidance formula, notably due to the presence of dissipative phenomena.)

When the velocity of the concretion obeys the guidance formula, the amplitude of the transverse wave at the location of the concretion is such that: $\vec{\nabla}{F}(\vec{\xi},t)=\vec{0}$ and $\frac{\partial F}{\partial t}(\vec{\xi},t)=0$.~\footnote{Eq.~(\ref{eth_symb}) becomes in the low-velocity approximation, $\vec{v}\,\psi(\vec{\xi},t) = \frac{c_m^2}{\ii\,\Omega_m}\vec{\nabla}\psi(\vec{\xi},t)$, whose real part yields $\vec{\nabla}{F}(\vec{\xi},t)=\vec{0}$. Put into the real part of Eq~(\ref{eth_symb}), we get $\frac{\partial F}{\partial t}(\vec{\xi},t)=0$.} Hence, the amplitude of transverse oscillation of the concretion, called $F_c$ in the following, remains constant in time. Moreover the concretion is located at a local extremum of the vibration amplitude field, $F(\vec{r},t)$. To give an image, apart from transverse oscillations, the concretion moves as if it surfs on the wave.\\

We have now to take into account the equation of motion for the concretion. When the concretion is not the source of the wave, the wave obeys the Klein-Gordon-like equation. The second term of~(\ref{eth_mvt}) is thus zero. This implies two possible behaviours when the $\varphi$-guidance formula is also satisfied. ($i$) The vector velocity of the concretion is constant. In other words, the concretion has in the reference frame $\mathcal{R}$ a uniform and linear motion. The velocity is thus given by the relation~(\ref{eth_symb}). ($ii$) The second possible behaviour can have a vector velocity of the concretion not constant in time. This solution occurs when the term $\frac{m_0}{2}(\partial_\mu \varphi\partial^\mu\varphi - \frac{\Omega_m^2}{c_m^2}\,\varphi^2)$ before the 4-velocity in the equation of motion is zero. Nevertheless this condition seems to be too demanding. In addition to the low-velocity approximation, we assume in this article that the time period of transverse oscillations is much shorter to the characteristic evolution time of the (perceivable) motion of the concretion. (This assumption is experimentally realised for bouncing droplets experiments and also for the bead sliding on a vibrating string~\cite{yc_sao1999}.) We verify with an example later (cf. Section~\ref{sssim_spher}) that the time period of a circular motion for the concretion in a spherical cavity ({\it i.e.} the characteristic evolution time of its velocity) is much longer than the time period of a transverse oscillation (whose order of magnitude is $1/\Omega_m$). Hence, when
\be \label{eth_symb_mvt}
\langle \: \left(\partial_\mu \varphi\partial^\mu\varphi - \frac{\Omega_m^2}{c_m^2}\,\varphi^2 \right)_{\vec{\xi}(t)} \: \rangle = 0 \,,
\ee
where $\langle (\cdots)_{\vec{\xi}(t)} \rangle$ denotes the time-averaged value at the location of the concretion, the motion of the concretion (in symbiosis with the wave) is only guided by the wave equation, more precisely by Eq.~(\ref{eth_symb}). The equation of motion for the concretion~(\ref{eth_mvt}) does not play any role on its movement (averaged during one oscillation). Note that, contrarily to the system proposed in this paper, the bead oscillator in~\cite{masselotte} cannot realise the `cancellation' of its equation of motion; this is a major difference between these two systems. The possible cancellation of the equation of motion in the symbiosis state explains why the velocity of the concretion is directly proportional to the gradient of the wave (due to the $\varphi$-guidance formula). This drastically contrasts with a movement derived from an equation of motion, in which the external influence is proportional to the acceleration.

When $\vec{v}$ satisfies the $\varphi$-guidance formula, $\partial_\mu \varphi\partial^\mu\varphi$ becomes $\frac{1}{\gamma_m^2\,c_m^2}(\frac{\partial \varphi}{\partial t})^2 = \frac{1}{c_m^2}(\frac{\dd \varphi}{\dd \tau})^2$, where $\frac{\dd \varphi}{\dd \tau}$ denotes the {\it particle velocity} of the concretion expressed in its proper time.~\footnote{In any reference frame where the concretion has the velocity $\vec{v}$, the particle derivative is written as $\frac{\dd }{\dd t}= \frac{\partial }{\partial t} + \vec{v}\cdot\vec{\nabla}$. \label{fnote}} Then, Eq.~(\ref{eth_symb_mvt}) becomes $\langle (\frac{\dd \varphi}{\dd \tau})^2\,  \rangle  = \langle  \Omega_m^2\,\varphi^2(\vec{\xi},t) \rangle$. This means that the concretion harmonically oscillates in its proper reference frame at the pulsation $\Omega_m$. (This also implies, as seen in the low-velocity approximation, that the amplitude of transverse oscillation of the concretion, $F_c$, remains constant in time.) Finally and more generally, when the concretion is in symbiosis and Eq.~(\ref{eth_symb_mvt}) is satisfied, we call this state in the following as a {\it reference state} for the concretion.\\

To sum up, we have analysed the special case where the concretion is not the source of the wave (and where there is no external potential). This state, called symbiosis, means that there is a strong relationship between the wave and the concretion. In particular, there is no longer feedback of the concretion on the transverse wave $\varphi$. When the symbiosis state is reached, the system is such that: ($i$) the wave $\varphi$ obeys a (homogeneous) Klein-Gordon-like equation and ($ii$) the symbiosis equation~(\ref{eth_ursymb}) is satisfied. The latter relation (in addition to the mass conservation of the concretion) implies that the speed of the concretion is constant in time and the concretion is governed by a general and covariant guidance formula~(\ref{eth_symb}), which leads in the low-velocity approximation to Eq.~(\ref{eth_vGuid}): an equivalent of the de Broglie guidance formula. The velocity of the concretion is thus directly proportional to the slope of the transverse wave at the location of the concretion. The equation of motion for the concretion~(\ref{eth_mvt}), whose right-hand side is zero in the symbiosis state, should not play any role. This occurs when that ($i$) the concretion has a linear and uniform movement or ($ii$) the concretion -- piloted by the guidance formula -- has a transverse oscillation at the pulsation $\Omega_m$ in its proper reference frame ({\it i.e.} the relation~(\ref{eth_symb_mvt}) is satisfied).

Therefore, the guidance formula for the concretion results from a situation in which the concretion is not the wave source any longer, {\it i.e.} there is in this state no feedback of the concretion on the wave. This means in our suggested system that the concretion is located at a local extremum of the transverse wave in its proper reference frame ({\it i.e.} $\vec{\nabla\varphi}(\vec{\xi},t)=0$ in $\mathcal{R}_0$). It is worth noting in a quantum viewpoint (cf.~\cite{holland1,holland2,holland3}) that the de Broglie guidance formula is also derived when there is no back-reaction of the particle on the wave. We also note that the interaction between the wave and the peaked quantum soliton~\cite{durt_2} leads the latter to follow a very close equation to the de Broglie guidance one.

Before investigating questions about energy, let us now mention the following point -- which is out of the scope of this article. The `cancellation' of the equation of motion mentioned above is only true in average, during one oscillation. It is thus conceivable that the concretion has a very rapid motion around its average location.

%%%%%%%%%%%%%
%%%%%%%%%%%%%
%%%%%%%%%%%%%
\subsection{Energetic considerations \label{ssth_E}}
When the concretions satisfies the guidance formula, it moves as if exists a {\it wave potential}, $Q$: an equivalent to the {\it Quantum Potential} established by de Broglie (cf.~\cite{ldb_tentative} \S \cRM{10}) and Bohm~\cite{bohm1}. In other words, the concretion moves under the influence of the potential $Q$ ({\it i.e.} $m_0\,\frac{\dd \vec{v}}{\dd t}=-\vec{\nabla}Q$), such that
\be \label{eth_Q}
Q = -\,\frac{m_0}{2} \, \left(\frac{c_m^2}{\Omega_m}\right)^2 \; \frac{\Delta F}{F} \,,
\ee
where $\Delta$ denotes the Laplace operator. This relation comes from the guidance formula~(\ref{eth_vGuid}) and the homogeneous wave equation~(\ref{eth_ch}) (see Appendix~\ref{sa_WPotential} for more details).\\

Now, we come back to the concretion defined as a high mass density of elastic medium in order evaluate its energy, $\Wconc$, and its linear momentum, $\vpconc$. They are defined from the time-averaged value of their corresponding density in the elastic medium. Note that energy and momentum densities derive from the Lagrangian density~(\ref{eth_l}), by means of the stress-energy tensor. This calculation is performed when the velocity of the concretion is governed by the $\varphi$-guidance formula~(\ref{eth_symb}). Calculations are made in a reference frame, $\mathcal{R}$, where the velocity of the concretion is $\vec{v}$. (See Appendix~\ref{sa_Econcretion} for details of calculation throughout this section.)

First, in addition to the $\varphi$-guidance formula, we assume that condition~(\ref{eth_symb_mvt}) is also satisfied; the concretion is then in its reference state. (Recall in this case that the concretion harmonically oscillates in its proper reference frame at the pulsation $\Omega_m$ with an amplitude $F_c$, and its motion, in symbiosis, can be non-rectilinear.) In this case, the energy and momentum of the concretion are written as
\bea \label{eth_Econc0}
 \Wconc & = & \gamma_m \: m_0\, \left[\frac{1}{2}\,\Omega_m^2\: F_c^2\right] \\
 \vpconc & = & \gamma_m \: m_0\, \frac{\left[\frac{1}{2}\,\Omega_m^2\: F_c^2\right]}{c_m^2} \; \vec{v} \,,
\eea
in which $\frac{1}{2}F_c^2 = \langle \varphi^2(\vec{\xi},t)\rangle$. We could be very tempted to set the experimental values in such a way that: 
\be \label{eth_Fc}
\frac{1}{2}\,\Omega_m^2\,F_c^2 = c_m^2 \,.
\ee 
By using this relation we retrieve, naturally, the usual equivalent formulas in relativity for the energy and momentum of a free mass $m_0$, namely $\Wconc = \gamma_m \,m_0\,c_m^2$ and $\vpconc=\gamma_m \,m_0\,\vec{v}$. We assume in the following that this condition is always fulfilled in the thought-experiment presented in this paper. (We can thus interpret in this system, for example, $\frac{1}{2}\,m_0\,v^2$ as a common kinetic energy of a mass $m_0$ in the low-velocity approximation.) But it could also be interesting to notice that condition~(\ref{eth_Fc}), expressed in $\mathcal{R}_0$, corresponds to: ($i$) an average quadratic transverse oscillation velocity of the concretion equal to $c_m^2$ ({\it i.e.} $\langle (\frac{\dd \varphi}{\dd \tau})^2\,  \rangle  = c_m^2$) and ($ii$) an energy density, in the very close neighbourhood to the concretion, equal to $\mathcal{T}$. (An experimenter could take advantage of these properties, maybe by means of a kind of self-adaptation.)

Second, we consider the general case where the wave $\varphi$ oscillates in $\mathcal{R}$ at the pulsation $\Omega=\Omega_m+\omega$, and the velocity of the concretion again obeys the $\varphi$-guidance formula. In the low-velocity approximation we get:
\begin{eqnarray}
 \Wconc & = & m_0\, c_m^2 \:+\: \frac{m_0\,c_m^2}{\Omega_m}\;\omega  \,,  \label{eth_Econc} \\
 \vpconc & = & m_0\: \vec{v} \label{eth_Pconc} \,.
\end{eqnarray}
The momentum of the concretion has the same expression as in classical mechanics. The expression about energy is particularly nice. This expression is maybe the most beautiful one in this article and perhaps brings the most novelties. The energy of the concretion encapsulates two terms: in one hand, the equivalent rest mass energy (like in relativity) and, on the other hand, an additional energy equal to a coefficient times the additive pulsation $\omega$. We call in the following $\Econc$ this additional energy, {\it i.e.} $\Econc = \frac{m_0\,c_m^2}{\Omega_m}\;\omega$. (We notice that the coefficient $\frac{m_0\,c_m^2}{\Omega_m}$, denoted later as $\hbe$, naturally appears from this calculation and is also derived from condition~(\ref{eth_Fc}) concerning the transverse oscillation amplitude of the concretion.) $\Econc$ looks like the energy of the system in quantum mechanics. Lastly, it is interesting to notice that the expression~(\ref{eth_Econc}) -- which comes from the density energy of the elastic medium -- is not common in the relativity point of view. For example, the additional energy can exist even if the concretion has no velocity, {\it i.e.} no kinetic energy in the reference frame.

Let us now shortly evoke that energy and momentum are also investigated in the quantum system~\cite{holland3}, but they have the same expressions as in common quantum mechanics, without including relativistic-like expressions. More interesting for our study, the peaked quantum soliton in~\cite{durt_1,durt_2} has an energy of the order $-mc^2$, {\it i.e.} an energy $mc^2$ would be required to destabilise it. (This should enable to connect the Durt's approach with the toy system presented in this paper, but it is not our scope to study this connection in depth here.)\\

What precedes, in particular the guidance formula, the wave potential $Q$ and the additional energy of the concretion, encourages us to go deeper into analogies between our toy model and quantum systems.

%%%%%%%%%%%%%%%%%%%%%%%%%%
%%%%%%%%%%%%%%%%%%%%%%%%%%
%%%%%%%%%%%%%%%%%%%%%%%%%%
\section{Quantum similarities}%\label{ssim}
In this section we first write general equations of the wave and of the concretion, in which quantum equivalences are more evident. Then we study equivalent systems which are commonly considered in quantum mechanics handbooks, where the system is in a stationary state and there is no external potential acting on the concretion and/or the wave.

Along this section we consider that ($i$) the velocity of the concretion is much lower than $c_m$ and ($ii$) the wave and the concretion are in symbiosis. In other words, calculations are made under the low-velocity approximation, the wave $\varphi$ obeys the (homogeneous) Klein-Gordon-like equation~(\ref{eth_ch}) and the velocity of the concretion is given by a guidance formula (\ref{eth_symb}) or (\ref{eth_vGuid}).

%%%%%%%%%%%%%
%%%%%%%%%%%%%
%%%%%%%%%%%%%
\subsection{Equivalent quantum equations \label{sssim_equa}}
As was shown in~\cite{masselotte}, in the low-velocity approximation the Klein-Gordon-like equation without source leads to 
\be \label{esim_schro}
\ii \,\frac{\partial \psi}{\partial t} = -\,\frac{c_m^2}{2\,\Omega_m}\,\Delta\psi \,.
\ee
(Appendix~\ref{sa_KG-Schro} gives a sketch of this derivation; and recall in this paper that there is no external potential acting on the concretion and/or the wave.) It is important to note that the Klein-Gordon-like equation~(\ref{eth_ch}) deals with the (real-valued) transverse wave, $\varphi$, while Eq.~(\ref{esim_schro}) deals with the (complex-valued) modulating wave, $\psi$.

A proportional coefficient between wave characteristics and particle characteristics was introduced in~\cite{masselotte}, namely:
\be \label{esim_hbe}
\hbe=\frac{m_0\,c_m^2}{\Omega_m} \,.
\ee
Note that this coefficient is not proper to the elastic medium, since it depends not only on characteristics of the medium but also on the mass $m_0$ of the concretion used in our thought-experiment. Nevertheless the coefficient $\hbe$, specific to the studied system, is very convenient to make a correspondence between wave and particle characteristics. Moreover this allows us to write some relations in our suggested system with the same form as the ones in equivalent quantum systems. For examples, using $\hbe$: Eq.~(\ref{esim_schro}) becomes strictly equivalent to the free Schrödinger equation, Eq.~(\ref{eth_vGuid}) to the de Broglie guidance formula and Eq.~(\ref{eth_Q}) to the quantum potential.

Let $\omega$ the additional pulsation (on $\Omega_m$) of the wave $\varphi$. The additional pulsation $\omega$ is contained in the modulating wave $\psi$, such that $\frac{\partial \Phi}{\partial t}=-\omega$ (cf. Eqs.~(\ref{eth_psi_def}) and (\ref{eth_psi_FPhi})). Hence, using Eq.~(\ref{esa_dPhi_dt}), the wave potential~(\ref{eth_Q}), the guidance formula~(\ref{eth_vGuid}) and $\hbe$ yield:
\be \label{esim_E_vQ}
\hbe \: \omega = \frac{1}{2}\,m_0\,v^2 \:+\: Q(\vec{\xi},t) \,.
\ee
An equivalent of this equation provides a way to evaluate the energy of a quantum system from a particle point of view (see {\it e.g.} \cite{Q_th_motion} \S 3.5).\\

It is particularly interesting to discuss the energy and momentum of the concretion written with $\psi$ in the light of quantum mechanics. From Eqs.~(\ref{eth_Econc}), (\ref{eth_Pconc}), the guidance formula~(\ref{eth_vGuid}) and $\hbe$, we get:
\begin{eqnarray}
 & & \Econc = \hbe \: \omega \,, \label{esim_Econc}\\
 & & \vpconc = \hbe\,\vec{\nabla}\Phi(\vec{\xi},t) \,. \label{esim_Pconc}
\end{eqnarray}
These relations have very close similarities with the ones in quantum mechanics, for a system without external potential. It is moreover interesting to notice that $\Econc$ and $\vpconc$ derive only from the phase $\Phi$ -- at the location of the concretion. But before to discuss them, let us write another usual form encountered in quantum mechanics. Since at the location of the concretion, the amplitude of vibration $F$ is constant in time and has a local extremum ({\it i.e.} $\frac{\partial F}{\partial t}(\vec{\xi},t)=0$ and $\vec{\nabla}F(\vec{\xi},t)=\vec{0}$), it follows:
\bea \label{esim_E_p_psi}
 & & \ii \,\hbe\: \frac{\partial \psi}{\partial t}(\vec{\xi},t)\, = \, \Econc \:\psi(\vec{\xi},t) \\
 & & \frac{\hbe}{\ii}\; \vec{\nabla}\psi(\vec{\xi},t)\, = \, \vpconc\: \psi(\vec{\xi},t) \,.
\eea 
One more time these equations have a direct counterpart in quantum mechanics; apart from they specifically concern the modulating wave $\psi$ at the location of the concretion. There is another difference: in quantum mechanics these equations deal with energy and momentum of the system, while here they concern a more concrete object, the concretion.

It thus appears that $\Econc$ and $\vpconc$ are (also) the energy and momentum of an equivalent quantum system.~\footnote{The rest mass energy has no importance in quantum mechanics {\it à la} Schrödinger.} In other words, the additional energy and the momentum of the concretion encapsulate ({\it i.e.} are in exact agreement with) the energy and momentum of an equivalent quantum system. This point confirms in our system what de Broglie had suggested, here restricted to energy and momentum: the particle accounts for quantities commonly attributed to the wave-like nature of the system in quantum mechanics (see {\it e.g.} \cite{ldb_tentative}, \S \cRM{11}). Lastly, it seems to us that the energy and momentum of the concretion -- which derive from the energy and momentum density in the elastic medium -- are more concrete, and very probably more easily measurable by an experimenter, than for example the (convenient) wave potential $Q$, the effective kinetic energy~\cite{masselotte} -- and even more the ``wave energy'' in bouncing droplets experiments~\cite{wave-mediated}.

%%%%%%%%%%%%%
%%%%%%%%%%%%%
%%%%%%%%%%%%%
\subsection{Free concretion \label{sssim_free}}%
In this section we study plane wave solutions for the concretion. According to the equivalent Schrödinger equation~(\ref{esim_schro}), the modulating wave is written as: $\psi=A\,\e^{\ii(k\,x-\omega\,t)}$, where $\omega=\frac{k^2\,c_m^2}{2\,\Omega_m}$ and $A$ denotes an amplitude of vibration. 

The concretion is located at a local extremum of the vibration amplitude field. In the plane wave case, the concretion can be located anywhere. The amplitude of oscillation $A=F_c$ should be given by Eq.~(\ref{eth_Fc}). The velocity of the concretion, given from the guidance formula, is $\vec{v} = \frac{c^2}{\Omega_m}\,k\,\vec{e}_x$, where $\vec{e}_x$ denotes the unit vector along the $(Ox)$ axis. Since $\vec{v}$ is constant in time, the equation of motion for the concretion is also satisfied. The moment of the concretion~(\ref{esim_Pconc}) is $\vpconc=\hbe\,k\,\vec{e}_x$. The additional energy of the concretion is thus equal to the expected value for a free mass $m_0$, namely its kinetic energy.

Let us shortly generalise the previous results for higher velocities. Let a concretion in symbiosis with a plane wave $\varphi = A\, \cos(k\,x-\Omega\,t)$. The Klein-Gordon-like equation~(\ref{eth_ch}) leads to $c_m^2\,k^2=\Omega^2-\Omega_m^2$ and the $\varphi$-guidance formula~(\ref{eth_symb}) (not restricted to the low-velocity limit) $\vec{v}=\frac{c_m^2 \, k}{\Omega}\, \vec{e}_x$; thus $\Omega=\gamma_m\,\Omega_m$. The concretion is then in its reference state. Since the concretion oscillates at $\Omega_m$ in its proper reference frame and provided condition~(\ref{eth_Fc}) is fulfilled, Eqs.~(\ref{eth_Econc0}) lead to: $\Econc=\gamma_m\,m_0\,c_m^2$ and $\vpconc=\gamma_m\,m_0\,\vec{v}$, as expected for a free mass satisfying equivalent relativistic equations.

It is maybe interesting to notice that the wave $\varphi$, with a progressive and plane phase and with a travelling and localised amplitude, given in~\cite{masselotte} (cf. Eq.~(18)), does probably not correspond to a true solution -- even if $\varphi$ obeys the Klein-Gordon-like (or the equivalent Schrödinger equation in the low-velocity approximation). Energy and momentum of the concretion are then different from the expected ones of a free mass. (This difference results from the fact that the concretion oscillates in its proper reference frame at a pulsation different from $\Omega_m$.)

%%%%%%%%%%%%%
%%%%%%%%%%%%%
%%%%%%%%%%%%%
\subsection{Concretion in a linear cavity \label{sssim_lin}}
We are looking for the behaviour of the concretion put into a linear cavity of length $L$, when the wave is standing in the laboratory reference frame. We consider the case of boundary conditions such that the wave amplitude is zero in $x=0$ and $x=L$. The aim of this section is to illustrate the general results seen in Section~\ref{sssim_equa}, to discuss the energy of the concretion and the superposition of eigenstates with a simple example. Calculations for a 3D rectangular box are straightforward and not made here.

From the equivalent Schrödinger equation and taking into account boundary conditions, the modulating wave is written as: $\psi(x,\,t) = A_0\, \sin(K_n\,x)\,\e^{-\ii\,\omega_n\,t}$, where $K_n = \frac{n\,\pi}{L}$ ($n$ being a natural number), $\omega_n = \frac{K_n^2\,c_m^2}{2\,\Omega_m}$ and $A_0$ denotes the maximum amplitude of $\psi$. The concretion is located at a local extremum of the vibration amplitude field, $F(x)=\sin(K_n\,x)$. According to the guidance formula, the velocity of the concretion is zero -- which directly implies that the equation of motion for the concretion is also satisfied. The maximum amplitude of vibration, $A_0$, is thus equal to the oscillation amplitude of the concretion, $F_c$, given by Eq.~(\ref{eth_Fc}). $\Econc$ is equal to the wave potential $Q$~(\ref{eth_Q}), here $\frac{m_0}{2}\frac{c_m^4}{\Omega_m^2}\,K_n^2$, in agreement with equivalent quantum systems. It is interesting to notice that, contrarily to the system with a bead oscillator~\cite{masselotte}, all the stationary solutions written above can here exist. Compared to the bead oscillator system, this is a major difference -- and also a substantial improvement.\\

The linear cavity example sheds light on the peculiar form of the additional energy of the concretion and, more precisely, its distinctive feature with regard to the one in relativistic and classical mechanics. To begin, it is worth noting that the concretion is not in its reference state (as it oscillates at $\Omega_m + \omega_n$ in its reference frame). The momentum $\vpconc$ of the concretion is here zero. Consequently, the kinetic energy in relativistic and classical mechanics of a mass $m_0$ should also be zero. But here, for the concretion, its additional energy, $\Econc$, is exactly equal to $\hbe\,\omega$ -- like in equivalent quantum systems. We can interpret this result as follows: The concretion is `more' than an usual point mass $m_0$ as in macroscopic systems. The concretion is not only a mass $m_0$, but has also specific properties due to its elastic-medium nature (see also Appendix~\ref{sa_comparison}). Since the elastic medium has a transverse vibration with an additional pulsation $\omega$, the concretion also oscillates with this additional pulsation. Consequently increases $\Econc$. In addition, it is interesting to notice that the increased energy of the concretion does not, in turn, imply that the concretion becomes a source of the wave. (This should be false with the bead oscillator~\cite{masselotte}.) The concretion remains in symbiosis with the wave. This point exemplifies the dual wave-particle nature of the concretion. The concretion, a high elastic medium density of mass $m_0$, behaves as the vibrating elastic medium around it.\\

Appendix~\ref{sa_superpos} deals with the superposition of eigenstates in the linear cavity. The total energy of the concretion seems to differ from the one expected from quantum mechanics. To overcome this problem, additional consideration would be required, in particular taking into account that the concretion could be a simplified representation of a soliton.

%%%%%%%%%%%%%
%%%%%%%%%%%%%
%%%%%%%%%%%%%
\subsection{Concretion in a spherical cavity \label{sssim_spher}}%
We are looking for the behaviour of the concretion put into a spherical cavity of radius $R$, such that the wave is zero at the boundaries. Note the wave $\varphi$ is not necessary standing in the laboratory reference frame -- contrarily to the one in~\cite{masselotte} -- but stationary in the sense of quantum mechanics.

Taking into account boundary conditions and using the equivalent Schrödinger equation, the modulating wave with spherical coordinates is written as: $\psi=F(r,\,\theta)\,\e^{\ii(m\,\phi\,-\,\omega_{n,\ell}\,t)}$, where $F(r,\,\theta) = A\: \mathrm{j}_\ell(K_{n,\ell}\,r) \: P_\ell^{m}(\cos \theta)$, $\omega_{n,\ell} = \frac{K_{n,\ell}^2\,c_m^2}{2\,\Omega_m}$, $\mathrm{j}_\ell$ denotes the spherical Bessel function of the first kind and order $\ell$, $P_\ell^m$ the associated Legendre polynomial (where $\ell$ and $m$ are two natural numbers such that $ |m| \leq \ell$) and $A$ an amplitude of vibration. Boundary conditions are satisfied by the fact that $K_{n,\ell}R$ is equal to the $n$-th zero of $\mathrm{j}_\ell(X)$. (Note that $\psi$ can also be expressed with spherical harmonics $Y_\ell^m(\theta,\,\phi)\propto P_\ell^m(\cos\theta)\,\e^{\ii m\phi}$ and, moreover, the above expression of $\psi$ is commonly established in other branches of physics, as in acoustics for example. The above expression is equivalent to the one for a quantum particle in a spherical cavity (see {\it e.g.}~\cite{landau_MQ} \S 33). (Calculations for a 2D circular cavity are straightforward and solutions use Bessel functions of the first kind.) Lastly, the solution can be degenerate since it allows the existence or the superposition of several solutions with different $m$ for the same $\omega_{n,\ell}$; for instance the superposition of two solutions, one with $m$ and the other one with $-m$, leads to a standing wave $\varphi$. Nevertheless, in the following we study the system when $\varphi$ has only one of the solution written above.

The concretion is located at a local extremum of the vibration amplitude field, $F$, which provides possible values of $r_c$ and $\theta_c$. The index $c$ denotes the location of the concretion. The velocity of the concretion, derived from the guidance formula~(\ref{eth_vGuid}), is written as $\vec{v}=m\,\frac{c_m^2}{\Omega_m\,r_c\,\sin \theta_c}\,\vec{e}_\phi$. Thus, the momentum of the concretion~(\ref{eth_Pconc}) is $\vpconc=\frac{m\,\hbe}{r_c\,\sin \theta_c}\,\vec{e}_\phi$. The wave potential $Q$~(\ref{eth_Q}) is here equal to $\frac{m_0}{2}\frac{c_m^4}{\Omega_m^2}(K_{n,\ell}^2-\frac{m^2}{r_c^2\,\sin^2\theta_c})$, in agreement with Eq.~(\ref{esim_E_vQ}). Lastly, as evoked for Eq.~(\ref{eth_symb_mvt}), we verify that the time period of a circular motion for the concretion is much longer greater than the time period of its transverse oscillation (equal to $\frac{2\pi}{\Omega_m+\omega_{n,\ell}}$, whose order of magnitude is $1/\Omega_m$ in the low-velocity approximation).~\footnote{Let $\omega_{\mathrm{circ}, v} = \frac{v}{r_c\,\sin \theta_c}$ the angular velocity of the concretion. The velocity of the concretion derived from the guidance formula yields $\omega_{\mathrm{circ}, v} = \frac{v^2}{c_m^2}\frac{\Omega_m}{m}$. This leads in the low-velocity approximation (note $v = \omega_{\mathrm{circ}, v} = 0$ when $m= 0$) to $\omega_{\mathrm{circ}, v}\ll \Omega_m$.}

Similarly to the energy and momentum of the concretion, the angular momentum of the concretion, $\vec{L}_\mathrm{conc}$, is defined from its corresponding density in the elastic medium. The angular momentum density derives from the stress-energy tensor of the system.~\footnote{The angular momentum density derives from the stress-energy tensor. Spatial integration around the location of the concretion and using time-averaged values (in the same manner as for $\Wconc$ and $\vpconc$) lead to $\vec{L}_\mathrm{conc}$.} In the low-velocity approximation: $\vec{L}_\mathrm{conc}=\vec{r}_c \times \vpconc$, where $\vpconc$ is given by Eq.~(\ref{eth_Pconc}). According to the expression of $\vec{v}$ given above, the projection of the angular momentum of the concretion along the $(Oz)$ axis is: $L_{\mathrm{conc},\,z}=m\,\hbe$. This value is equal to the one commonly allocated to the system or the wave in the equivalent quantum system -- as for energy and linear momentum values. (At this step we cannot evaluate the quadratic angular momentum of the concretion from the corresponding density value in the elastic medium. But let us just evoke that the value of $\vec{L}^2$ in quantum mechanics, with a particle point of view, necessitates to add something else to $(m_0\,\vec{r}\times\vec{v})^2$, like the quantum potential $Q$ to $\frac{1}{2}\,m_0\,v^2$ for the energy (see {\it e.g.}~\cite{Q_th_motion} \S 3.5). Lastly, let us recall that by using the {\it effective velocity}~\cite{masselotte} of the concretion we very easily retrieve the value of $\vec{L}^2$.) \\

We take advantage of this study to investigate the influence of the equation of motion on the movement of the concretion (derived from the guidance formula). The term in brackets in the left hand side of Eq.~(\ref{eth_mvt}), in the low-velocity approximation and averaged during one oscillation time, has the following form, very general and not restricted to the spherical cavity system: $\frac{\dd }{\dd t}[\frac{Q}{c_m^2}\,\vec{v}]$.~\footnote{$\langle \partial_\mu \varphi \partial^\mu \varphi \rangle = (\frac{(\Omega_m + \omega)^2}{c_m^2}-\frac{v^2}{c_m^4}(\Omega_m+\omega)^2)\,\langle\varphi^2\rangle$, by using the guidance formula~(\ref{eth_symb}) and $\frac{\partial F}{\partial t}=0$ at the location of the concretion. Next, the low-velocity approximation (related to $\omega\ll\Omega_m$), Eq.~(\ref{esim_E_vQ}) and condition~(\ref{eth_Fc}) are used.} When the concretion is in symbiosis with the wave, the right hand side of the equation of motion is zero, hence $\frac{\dd }{\dd t}[\frac{Q}{c_m^2}\,\vec{v}]=\vec{0}$. Recall that the guidance formula is also written as $\frac{\dd }{\dd t}[m_0\,\vec{v}]=-\vec{\nabla}Q$. The equation of motion is not in agreement with the guidance formula for the concretion in a cavity, but its influence is very small. Indeed, $|\frac{Q}{c_m^2}|\ll m_0$, since $Q$ -- which has the same order of magnitude as the additional energy of the concretion -- is much lower than its rest mass energy in the low-velocity approximation. (When the effective mass in the equation of motion, $\frac{Q}{c_m^2}$, is zero, {\it i.e.} the concretion oscillates in its proper reference frame at the pulsation $\Omega_m$, the equation of motion has, in average, no influence.) A very small perturbation on the right side of the equation of motion could allow it to be in agreement with the velocity given by the guidance formula. In other words, the velocity of the concretion is mainly given by the guidance formula, and the equation of motion perturbs very weakly this velocity. This might imply an instability (which are out the scope of this paper) of the motion given by the guidance formula.

%%%%%%%%%%%%%%%%%%%%%%%%%%
%%%%%%%%%%%%%%%%%%%%%%%%%%
%%%%%%%%%%%%%%%%%%%%%%%%%%
\section*{Conclusion}
In this paper we have studied a theoretical and classical system mostly inspired by the bouncing droplets experiments (see~\cite{bush_review15} for a review) -- and desired to be implementable in practice. It consists of ($i$) an elastic medium, which carries transverse waves, $\varphi$, governed by a Klein-Gordon-like equation, and ($ii$) one high elastic medium density, considered as a point of mass $m_0$ and called concretion. The system is wave monistic. The nature of the concretion strengthens the dual wave-particle nature of the system in comparison to an (external) bead oscillator~\cite{masselotte}. The system is described by the Lagrangian density~(\ref{eth_l}), which is covariant with respect to the elastic medium having a propagation speed of the wave $c_m$. The Lagrangian density allows us to establish the equation of motion for the concretion~(\ref{eth_mvt}) and the wave equation~(\ref{eth_ch}). We have studied the special case in which the concretion is such that the wave source cancels. The concretion and the wave are in intimate harmony, called symbiosis, without back-reaction of the concretion on the wave. (In this case, an observer focusing on the wave would be blind to the concretion.) In this state in which the system is subsequently studied: ($i$) the wave obeys a (homogeneous) Klein-Gordon-like equation, leading to an equivalent free Schrödinger equation~(\ref{esim_schro}) for the modulation wave, $\psi$~(\ref{eth_psi_def}), in the low velocity approximation and ($ii$) the symbiosis equation~(\ref{eth_ursymb}). According to the conservation of the mass of the concretion, the latter equation implies that the speed of the concretion is constant (which is in agreement with the conservation of the energy a particle in the absence of external potential) and, in addition, the concretion is located at an extremum of the transverse wave $\varphi$ in its proper reference frame. (Apart from the transverse oscillation, the concretion moves as if it surfs on a wave.) This leads to the general and covariant guidance formula~(\ref{eth_symb}) -- expressed with the transverse displacement wave $\varphi$ -- and, in the low velocity approximation, to an equivalent de Broglie-Bohm guidance formula~(\ref{eth_vGuid}). Under this approximation the concretion moves also as if an equivalent quantum potential~(\ref{eth_Q}) exists. Lastly, the concretion is studied in a linear and a spherical cavity. Equivalent quantum stationary states are obtained.

The suggested system studied in this paper shows a transverse wave carried by an elastic medium which interacts with a peaked concentration of mass and energy (the high elastic medium density, called concretion). Thus it is reminiscent of the de Broglie double solution theory~\cite{ldb_tentative,ldb_interpretation1987}; but in our paper it is presented in a simplified version and in a classical system. We note that the proposed system retrieves equivalent quantum stationary states (without external potential~\footnote{A paper including an external potential is in preparation.}) when the concretion and the wave are in symbiosis. Moreover we note that the suggested system could also allow us to investigate phenomena out of the scope of quantum mechanics. For instance, when there is no longer symbiosis (in this case the guidance formula is not satisfied and $\psi$ does not obey the equivalent Schrödinger equation) and by mean of numerical simulations: transitions between two stationary states.

Let us shortly mention two points. ($i$) At this step, the mass $m_0$ of the concretion is arbitrary. In this paper there is no relation to assign its value from, for instance, the pulsation $\Omega_m$ or the tension $\mathcal{T}$ proper to the elastic medium. Additional considerations than the ones presented in this paper would be required. (For instance, considering the concretion as a soliton~\cite{durt_1,durt_2}.) ($ii$) This system exhibits one more kind of equation than in quantum mechanics: the equation of motion for the concretion. (All results described above derive from the wave equation.) When the concretion has a transverse oscillation in its proper reference frame at the pulsation $\Omega_m$ (the one in the Klein-Gordon-like equation), the equation of motion is `cancelled' in average, during one oscillation. When this condition is not satisfied and in the low-velocity approximation, the equation of motion should very weakly perturb, and maybe destabilise, the velocity and trajectory of the concretion given by the guidance formula.\\

However the more interesting seems to concern energy, $\Wconc$, and momentum, $\vpconc$, of the concretion. They naturally result from their corresponding densities in the oscillating elastic medium -- {\it i.e.} from the stress-energy tensor of the system. When the concretion is in its reference state ({\it i.e.} it is in symbiosis and has a transverse oscillation in its proper reference frame at the pulsation $\Omega_m$) and provided a condition about the oscillation amplitude of the concretion~(\ref{eth_Fc}) is fulfilled, $\Wconc$ and $\vpconc$ are written as the ones of a free mass $m_0$ in relativity -- with regard to the specific propagation speed of the wave, $c_m$, of the elastic medium --, more precisely $\Wconc = \gamma_m \,m_0\,c_m^2$ and $\vpconc=\gamma_m \,m_0\,\vec{v}$. This seems not very surprising, since the Lagrangian density of the system is covariant -- with respect to $c_m$. (This is due to the fact that a Klein-Gordon-like equation governs the wave dynamics in the elastic medium and this equation is invariant by the Lorentz-Poincaré transformation.) But more strikingly is the energy of the concretion when it is not necessarily in its reference state. When the wave $\varphi$ oscillates at the pulsation $\Omega_m + \omega$ in the laboratory reference frame and in the low-velocity approximation, we get:
\be
\Wconc = m_0\, c_m^2 \:+\: \hbe\,\omega \,,
\ee
where naturally the proportional coefficient $\hbe$ appears~(\ref{esim_hbe}). (This coefficient was previously introduced in this kind of system for a convenient reason, in order to ensure proportionality between wave characteristics and particle characteristics~\cite{masselotte}.) For us it seems particularly interesting to notice, in our suggested system, that both the equivalent rest mass energy ($m_0\,c_m^2$) and the coefficient $\hbe$ come naturally from the same condition of the transverse oscillation amplitude of the concretion~(\ref{eth_Fc}), {\it i.e.} $\langle\Omega_m^2\, \varphi^2(\vec{\xi},t)\rangle = c_m^2$. This condition, assumed here to be experimentally performed, does not seem too surrealistic, since this means in the proper reference frame of the concretion and when it is in its reference state that the average quadratic transverse oscillation velocity of the concretion is equal to $c_m^2$ and/or the energy density in the very close neighbourhood to the concretion is equal to the `tension' $\mathcal{T}$ proper to the elastic medium. Anyway, in our system the energy of the concretion encapsulates two kinds of energies, which have a counterpart in two different branches of physics: relativity (with the equivalent rest mass energy, $m_0\,c_m^2$) and quantum mechanics (with the equivalent energy of the system, $\hbe\,\omega$, in the low-velocity approximation).

In the toy and classical system presented in this paper, it appears that the common (equivalent) relativistic expressions for energy could be naturally and easily extended, in particular when the concretion is not in its reference state, in the way that they include typically (equivalent) quantum energy. This extension for this suggested system, rooted in the dynamics of a high density of elastic medium embedded in the same elastic medium, does not seem particularly surprising in the end: this is also reminiscent of the original approach of Henri Poincaré about the theory of relativity~\cite{poincare_dyn_e-,pierseaux_RR}. If the toy system suggested in this paper would capture some features of the physical world, we expect that the theory of special relativity (at least for energetic considerations) as this is nowadays commonly used, will be naturally improved -- probably, and ironically, according to a viewpoint more in agreement with the Poincaré's one.

Now the focus is on equivalent quantum equations. The additional energy $\Econc$ (to the equivalent mass energy $m_0\,c_m^2$) of the concretion and $\vpconc$ were expressed with the modulating wave $\psi$. It appears (cf. Eqs.~(\ref{esim_Econc}) (\ref{esim_Pconc}) and (\ref{esim_E_p_psi})) that the equivalent additional energy and momentum of the concretion are in exact agreement with the equivalent quantum values. These latter are commonly assigned in quantum mechanics to the wave or to the system. Since $\Econc$ and $\vpconc$ concern a concrete and localised object (the concretion) and very probably seem easy to measure by an experimenter, we can suggest: measures of energy and momentum assigned to the system only concern the ones of the concretion. In other words, what precedes allows us to suppose (when this system will be experimentally performed, if possible) about energy and momentum: not only the concretion reflects the values assigned to the wave, but also measures are only related to the concretion.

%%%%%%%%%%%%%%%%%%%%%%%%%%
%%%%%%%%%%%%%%%%%%%%%%%%%%
%%%%%%%%%%%%%%%%%%%%%%%%%%
\subsubsection*{Aknowledgments}
We thank Hervé Mohrbach and Alain Bérard for help and discussions. We would like to thank the Reviewer for his constructive suggestions and insightful comments on the paper.

%%%%%%%%%%%%%%%%%%%%%%%%%%
%%%%%%%%%%%%%%%%%%%%%%%%%%
%%%%%%%%%%%%%%%%%%%%%%%%%%
%\appendix 
\section*{Appendices}

\renewcommand{\theequation}{A\arabic{equation}}
\setcounter{equation}{0} 

\renewcommand{\thesubsection}{A1}
\subsection{Lagrangian density of a system with a bead versus one with an elastic medium concretion and some consequences \label{sa_comparison}}
The system presented in~\cite{masselotte} consists of ($i$) a bead oscillator, {\it i.e.} a punctual mass with an ``internal clock'', by which the mass tends to oscillate at a proper pulsation $\Omega_0$ through a quadratic potential, and ($ii$) the same (homogeneous) elastic medium as here. The Lagrangian density of this system is thus written as
\be \label{esa_lbead}
\mathcal{L}_{\mathrm{syst.\, with\, bead}} = \frac{1}{2}\,\rho_0 \left[\left(\frac{\dd \varphi}{\dd \tau}\right)^2 \; -\; \Omega_0^2\,\varphi^2 \right]\, + \frac{1}{2}\, \mathcal{T}\,\left[\partial_\mu \varphi \, \partial^\mu \varphi  \,-\, \frac{\Omega_m^2}{c_m^2} \varphi^2 \right] \,,
\ee
where $\frac{\dd }{\dd \tau}$ is the particle derivative (cf. Footnote~\ref{fnote}) expressed with $\tau$ the natural time of the bead, $\tau$, and $\rho_0$ denotes its natural mass density. $\rho_0$ of a bead of mass $m_0$ is again written as in Eq.~(\ref{eth_rho0}) in the proper reference frame, $\mathcal{R}_0$, of the bead.

The difference between the Lagrangian density of a system with a bead oscillator~(\ref{esa_lbead}) and the one with a point-like high density medium~(\ref{eth_l}) naturally lies in the part dedicated to the `particle' -- {\it i.e.} the bead or the concretion. A basic comparison between the two Lagrangian densities emphasises the following points.
\bi 
\item[-] The system with a bead has a wave-particle duality nature, which is {\it de facto} imposed from the beginning -- from the density Lagrangian. In contrast, the system with a concretion is wave monistic. 
\item[-] The system with a bead has two reference pulsations, the one of the elastic medium or wave ($\Omega_m$) and the one proper to the bead oscillator ($\Omega_0$). Of course, the latter one does not exist in the monistic system with the concretion -- and only remains the reference pulsation of the wave, $\Omega_m$, which also describes the pulsation of the quadratic potential acting on the `particle' concretion.
\item[-] The bead involves in the Lagrangian density $\left(\frac{\dd \varphi}{\dd \tau}\right)^2$, while the concretion $c_m^2\,\partial_\mu \varphi \, \partial^\mu \varphi$ (recall in the Lagrangian density~(\ref{eth_l}) that $\mathcal{T}=\mu_0\,c_m^2$). For the sake of simplicity we place ourself in the proper reference frame $\mathcal{R}_0$ of the `particle', where the subscript $0$ means this reference frame. In one side $\left(\frac{\dd \varphi}{\dd \tau}\right)^2=\left(\frac{\partial \varphi}{\partial t_0}\right)^2$, while on the other side $c_m^2\,\partial_\mu \varphi \, \partial^\mu \varphi = \left(\frac{\partial \varphi}{\partial t_0}\right)^2 - c_m^2\,\left(\vec{\nabla}_0\varphi\right)^2$. The concretion behaves thus as the bead, but with one important difference: on the contrary to the bead, the concretion takes into account the wave slope (at its location).~\footnote{Note that this difference vanishes when the concretion is in symbiosis with the wave (cf. Section~\ref{ssth_guidance}).} This means that the concretion, due to its elastic medium nature, is `more' than a macroscopic point mass as the bead. 
\ei

The concretion is admittedly `more' than an usual point mass as in macroscopic systems, but it is associated for all practical purposes to a material point. The concretion, as a macroscopic point mass, has indeed well characterised location, velocity, energy and linear momentum (cf. Section~\ref{ssth_E}) and angular momentum (cf. Section~\ref{sssim_spher}). However, in comparison to the bead, the elastic-medium nature of the concretion provides some specificities as follows.
\bi 
\item[-] The concretion can only be piloted by the wave. On the contrary to the bead, the usual equation of motion of the concretion can be `cancelled' (under certain conditions), which leads to the concretion to be only piloted by the wave (cf. Section~\ref{ssth_guidance}).
\item[-] Conditions for which the `particle' is no longer a wave source are less drastic for the concretion than for the bead (cf. Section~\ref{ssth_guidance}). (It suffice for the concretion to be located at a local extremum of the transverse wave in its reference frame, whereas the bead must oscillate in its reference frame at its proper pulsation.) On the contrary to a system with a bead, a system with a concretion can have several stationary states as in equivalent quantum mechanics. 
\ei

%%%%%%%%%%%%%%%%%%%%%%%%%%
%%%%%%%%%%%%%%%%%%%%%%%%%%
%%%%%%%%%%%%%%%%%%%%%%%%%% 
\renewcommand{\thesubsection}{A2}
\subsection{Calculation of the wave equation and the equation of motion for the concretion \label{sa_equas}}
The equation of motion for the concretion comes from a principle of least action, when the four-position of the concretion is subjected to a small change, $\xi^\alpha \rightarrow \xi^\alpha + \delta\xi^\alpha$, while the wave field $\varphi$ is fixed. This is applied to the Lagrangian of the concretion $L_c$. From the Lagrangian density (\ref{eth_l}) and the mass density of the concretion (\ref{eth_rho0}), the Lagrangian of the concretion expressed in a reference frame where the concretion has a velocity $\vec{v}$ is written as 
\be 
L_c = \frac{m_0\,c_m^2}{2\,\gamma_m} \left( \partial_\mu \varphi\partial^\mu\varphi - \frac{\Omega_m^2}{c_m^2}\,\varphi^2 \right) \,,
\ee
where $\gamma_m=\frac{1}{\sqrt{1-v^2/c_m^2}}$. $L_c$ is here the only Lagrangian which depends on the location of the concretion.

$\xi^\alpha \rightarrow \xi^\alpha + \delta\xi^\alpha$ leads at first order to $\varphi(\xi^\alpha) \rightarrow \varphi(\xi^\alpha) + \partial_\alpha \varphi(\xi^\alpha) \: \delta\xi^\alpha$, $\partial_\mu\varphi \rightarrow \partial_\mu\varphi + \partial_\mu\left(\partial_\alpha\varphi \:\delta\xi^\alpha\right)$ and $\dd \tau = \sqrt{\dd \xi^\mu \dd \xi_\mu}/c \rightarrow \dd \tau + \delta(\dd \tau)$ where $\delta(\dd \tau) = \frac{U_\alpha}{c_m^2}\, \dd(\delta\xi^\alpha)$ and $U^\alpha=\frac{\dd \xi^\alpha}{\dd \tau}$. Recall that $\tau$ is the proper time of the concretion (so $\dd t = \gamma_m\,\dd \tau$). Thus, the conditions for which the small change $\delta\xi^\alpha$ implies at the first order $\delta(\int L_c\, \dd t) = 0$, where boundary values are fixed, are such that 
\be 
\int m_0\,c_m^2\,\partial_\mu\left(\partial_\alpha\varphi \:\delta\xi^\alpha\right)\partial^\mu\varphi \;\dd \tau - \int m_0\, \Omega_m^2\,\varphi\,\partial_\alpha\varphi \,\delta\xi^\alpha\;\dd \tau + \int \frac{m_0}{2}\left(\partial_\mu \varphi \, \partial^\mu \varphi - \frac{\Omega_m^2}{c_m^2} \varphi^2 \right)\: U_\alpha\frac{\dd(\delta\xi^\alpha)}{\dd \tau}\dd \tau = 0 \,. 
\ee
Integrating by parts the first and the third integrals, with fixed end points, and since the small change $\delta\xi^\alpha$ is arbitrary, lead to the equation of motion~(\ref{eth_mvt}).\\

The wave equation comes from a principle of least action, when the wave field is subjected to a small change, $\varphi \rightarrow \varphi + \delta\varphi$, while the four-position of the concretion is fixed. 

$\varphi \rightarrow \varphi + \delta\varphi$ leads to $\delta(\varphi^2) = 2\varphi\,\delta\varphi$ and $\delta(\partial_\mu \varphi\,\partial^\mu \varphi)=2\partial_\mu (\delta\varphi)\,\partial^\mu \varphi$. According to Eq.~(\ref{eth_l}) the conditions for which the small change $\delta \varphi$ implies at first order $\delta (\int \mathcal{L}\, \dd t\, \dd^3 \vec{r}) = 0$, where boundary values are fixed, are written as
\be 
\int \mathcal{T}\left(1+\frac{\rho_0(\vec{r},t)}{\mu_0}\right)\left[\partial_\mu (\delta\varphi)\,\partial^\mu \varphi - \frac{\Omega_m^2}{c_m^2}\,\varphi\,\delta\varphi \right] \dd t\, \dd^3 \vec{r} = 0 \,.
\ee
Integrating by parts the term with $\partial_\mu (\delta\varphi)$, with fixed end points, and since the small change $\delta\varphi$ is arbitrary, provide the wave equation~(\ref{eth_ch}). (Note that the generalised Euler-Lagrange equation leads to the same result, as expected.)

%%%%%%%%%%%%%%%%%%%%%%%%%%
%%%%%%%%%%%%%%%%%%%%%%%%%%
%%%%%%%%%%%%%%%%%%%%%%%%%%
\renewcommand{\thesubsection}{A3}
\subsection{Calculation of the wave potential \label{sa_WPotential}}
Calculations in this Appendix are near for example to the ones in~\cite{applied_bohmian}. The particle derivative of the guidance formula~(\ref{eth_vGuid}) is written as
\be 
\frac{\dd \vec{v}}{\dd t} = \frac{c_m^2}{\Omega_m}\left(\frac{\partial\: \vec{\nabla} \Phi}{\partial t} + (\vec{v}\cdot\vec{\nabla})\vec{\nabla}\Phi\right) \,.
\ee
Using Eq.~(\ref{eth_vGuid}) for the velocity and some basic vector calculus identities yield:
\be \label{esa_dv_dt}
\frac{\dd \vec{v}}{\dd t} = \frac{c_m^2}{\Omega_m} \: \vec{\nabla}\left(\frac{\partial\, \Phi}{\partial t} + \frac{c_m^2}{2\:\Omega_m} (\vec{\nabla} \Phi)^2 \right) \,.
\ee
The term in brackets in the right side is evaluated from the wave equation~(\ref{eth_ch}) without any source wave -- since the concretion and the wave are assumed in these calculations to be in symbiosis. By using the complex notation of $\varphi$ (\ref{eth_psi_def}) with $\psi = F\,\e^{\ii\,\Phi}$, the real part of Eq.~(\ref{eth_ch}) leads to:
\be \label{esa_dPhi_dt}
\frac{2\,\Omega_m}{c_m^2}\,\frac{\partial\, \Phi}{\partial t} \:F - \Delta F + (\vec{\nabla} \Phi)^2 \: F = 0 \,,
\ee
in which terms with $(\frac{\partial\, \Phi}{\partial t})^2$ and $\frac{\partial ^2 \, F}{\partial t^2}$ are neglected in the low-velocity approximation. Put into Eq.~(\ref{esa_dv_dt}) it appears that the concretion moves under the influence of the potential $Q$ given in Eq.~(\ref{eth_Q}).

%%%%%%%%%%%%%%%%%%%%%%%%%%
%%%%%%%%%%%%%%%%%%%%%%%%%%
%%%%%%%%%%%%%%%%%%%%%%%%%%
\renewcommand{\thesubsection}{A4}
\subsection{From the Klein-Gordon-like equation without source to the equivalent free Schrödinger equation \label{sa_KG-Schro}}
The derivation of the free Schrödinger from the Klein-Gordon equation (without source) in the low-velocity approximation is mentioned by de Broglie (see e.g. \cite{ldb_tentative} \S \cRM{2}.7) and is well-known in the literature (see e.g. \cite{zee} \S \cRM{3}.5). Here, by using the modulating wave $\psi$ contained in $\varphi$ (\ref{eth_psi_def}), the Klein-Gordon-like equation~(\ref{eth_ch}) without source yields
\be 
\mathrm{Re}\left[\left(\frac{1}{c_m^2}\left[-\,\Omega_m^2\,\psi - 2\,\ii\,\Omega_m\,\frac{\partial \psi}{\partial t} + \frac{\partial^2 \psi}{\partial t^2}\right] - \, \Delta\psi \, + \, \frac{\Omega_m^2}{c_m^2}\,\psi\right) \e^{-\ii\,\Omega_m t}\right] = 0 \,.
\ee
Since this equation holds for any $t$, we get:
%After elimination of $\e^{-\ii \Omega_m t}$, the latter equation becomes: 
\be \label{esa_KG-Schro}
- \frac{1}{c_m^2}\frac{\partial^2 \psi}{\partial t^2} - 2\,\ii\frac{\Omega_m}{c_m^2}\frac{\partial \psi}{\partial t}  - \Delta\psi = 0 \,.
\ee %in which $\e^{-\ii \Omega_m t}$ is eliminated.
In the low-velocity approximation, the term $\frac{\partial^2 \psi}{\partial t^2}$ is negligible with respect to the other terms. 

To convince ourselves of the validity of this approximation, we consider a free concretion in symbiosis with a plane wave $\varphi$ (cf. also Section~\ref{sssim_free}). By using $\varphi = A\, \cos(\vec{k}\cdot\vec{r}-\Omega\,t)$, the Klein-Gordon-like equation~(\ref{eth_ch}) leads to $c_m^2\,k^2=\Omega^2-\Omega_m^2$; and the $\varphi$-guidance formula~(\ref{eth_symb}) leads to $\vec{v}=\frac{c_m^2 \, \vec{k}}{\Omega}$. Thus $\Omega = \frac{1}{\sqrt{1-v^2/c_m^2}}\,\Omega_m$. According to Eq.~(\ref{eth_psi_def}), the modulating wave is written as $\psi = A\,\e^{\ii(\vec{k}\cdot\vec{r}-\omega\,t)}$, where $\omega = \Omega - \Omega_m$. This leads to $\omega=\frac{1}{2}\frac{v^2}{c_m^2}\,\Omega_m$ in the low-velocity approximation. Then, the term $\frac{1}{c_m^2}\frac{\partial^2 \psi}{\partial t^2}$ is proportional to $\frac{\Omega_m^2}{c_m^2}\frac{v^4}{c_m^4}\,\psi$, while the two other terms in Eq.~(\ref{esa_KG-Schro}) are proportional to $\frac{\Omega_m^2}{c_m^2}\frac{v^2}{c_m^2}\,\psi$. 

In the low-velocity approximation, the modulating wave $\psi$ is thus governed by the equivalent free Schrödinger equation~(\ref{esim_schro}).

%%%%%%%%%%%%%%%%%%%%%%%%%%
%%%%%%%%%%%%%%%%%%%%%%%%%%
%%%%%%%%%%%%%%%%%%%%%%%%%%
\renewcommand{\thesubsection}{A5}
\subsection{Energy and linear momentum of the concretion \label{sa_Econcretion}}
As usual, the energy density, $\rho_e$, and the momentum density, $\vec{g}$, are evaluated from the stress–energy tensor $T$ of the system ($T^{\mu\nu}=\frac{\partial \,\mathcal{L}}{\partial (\partial_\mu \varphi)}\,\partial^\nu\varphi - \eta^{\mu\nu}\,\mathcal{L}$, where $\eta^{\mu\nu}$ denotes the Minkowski metric, with the signature $(+,-,-,-)$ adopted throughout this article). The Lagrangian density~(\ref{eth_l}), where $\mathcal{T} = \mu_0\,c_m^2$, leads to
\bea \label{esa_rho_e}
& & \rho_e = \frac{\mu_0}{2}(1 + \frac{\rho_0}{\mu_0})\left[\left(\frac{\partial \varphi}{\partial t}\right)^2 + \left(c_m\,\vec{\nabla}\varphi\right)^2 + \Omega_m^2\,\varphi^2\right]  \\
& & \vec{g} = -\,\mu_0\,(1 + \frac{\rho_0}{\mu_0})\,\frac{\partial \varphi}{\partial t}\: \vec{\nabla}\varphi \,. 
\eea
In the following we use their time-averaged values during one oscillation (written as $\langle\cdots\rangle$). Spatial integration around the location of the concretion provides the energy, $\Wconc$, and the momentum, $\vpconc$, of the concretion. The following expressions are given in a reference frame, $\mathcal{R}$, where the concretion has the velocity $\vec{v}$. \\

We evaluate first the case for which condition~(\ref{eth_symb_mvt}) is satisfied, in addition to the $\varphi$-guidance formula. Eqs.~(\ref{eth_rho0}), (\ref{eth_symb}), (\ref{eth_symb_mvt}) and a little bit of algebra yield: 
\bea 
 \Wconc & = & \gamma_m \: m_0\: \Omega_m^2\: \langle\, \varphi^2(\vec{\xi},t)\, \rangle \\
 \vpconc & = & \gamma_m \: m_0\: \frac{\Omega_m^2}{c_m^2} \: \langle \, \varphi^2(\vec{\xi},t)\, \rangle \: \vec{v} \,.
\eea
We have notably used $\langle \frac{1}{\gamma_m^2} (\frac{\partial \varphi}{\partial t}(\vec{\xi},t))^2 \, \rangle = \Omega_m^2\,\langle \varphi^2(\vec{\xi},t) \rangle$. Since the amplitude of oscillation of the concretion, $F_c$, remains constant in time, it follows equations~(\ref{eth_Econc0}).\\

Now we consider that the concretion has just its velocity given by the $\varphi$-guidance formula~(\ref{eth_symb}). Let $\Omega$ (where $\Omega=\Omega_m+\omega$) the pulsation of the wave $\varphi$ in $\mathcal{R}$. In the same manner as above, Eq.~(\ref{esa_rho_e}) leads to

\bea \label{esa_Econc}
 \Wconc & = & \frac{1}{2}\,\frac{m_0}{\gamma_m}\:  \left[\left(1+\frac{v^2}{c_m^2}\right)\, \left\langle\, \left(\frac{\partial \varphi(\vec{\xi},t)}{\partial t}\right)^2\; \right\rangle \;+\; \Omega_m^2\, \langle\, \varphi^2(\vec{\xi},t)\, \rangle \right]   \\
\vpconc & = & \frac{m_0}{\gamma_m\, c_m^2}\: \left\langle\,\left(\frac{\partial \varphi(\vec{\xi},t)}{\partial t}\right)^2\; \right\rangle  \: \vec{v}\,.
\eea 
In the low-velocity approximation, $\frac{v^2}{c_m^2}$ and $\frac{\omega}{\Omega_m}$ have the same order of magnitude. (To convince us, consider how a pulsation $\Omega_m$ in $\mathcal{R}_0$ becomes in $\mathcal{R}$, for example by using Lorentz-Poincaré transformation.) Moreover $\langle\, (\frac{\partial \varphi}{\partial t}(\vec{\xi},t))^2 \, \rangle = \Omega^2\,\langle\, \varphi^2(\vec{\xi},t) \,\rangle$ -- because the magnitude $F$ at the location of the concretion is $F_c$, constant in time. Taking into account condition~(\ref{eth_Fc}), equations (\ref{esa_Econc}) become Eqs.~(\ref{eth_Econc}) and (\ref{eth_Pconc}) at first-order approximation in $\frac{v^2}{c_m^2}$ and $\frac{\omega}{\Omega_m}$.

%%%%%%%%%%%%%%%%%%%%%%%%%%
%%%%%%%%%%%%%%%%%%%%%%%%%%
%%%%%%%%%%%%%%%%%%%%%%%%%%
\renewcommand{\thesubsection}{A6}
\subsection{Superposition of eigenstates in the linear cavity \label{sa_superpos}}
We study the superposition of eigenstates for the concretion in the linear cavity (cf. Section~\ref{sssim_lin}). To have a deeper meaning, we deal with the transverse wave, $\varphi$, rather than the modulating wave, $\psi$.  According to the Klein-Gordon-like equation~(\ref{eth_ch}) without source and the boundary conditions, an eigenstate is written as $\varphi_n(x,t)=A_n\,\sin(K_n\,x)\,\cos(\Omega_n\,t -\theta_n)$, where $K_n = \frac{n\,\pi}{L}$,  $c_m^2\,K_n^2 = \Omega_n^2-\Omega_m^2$, $A_n$ denotes an amplitude of transverse oscillations and $\theta_n$ a phase shift, irrelevant in this study. (This expression is in agreement with the one of $\psi$ written in Section~\ref{sssim_lin} when $\omega_n(=\Omega_n-\Omega_m) \ll \Omega_m$.) For the sake of simplicity we consider two eigenstates, $n=1$ and $3$ (two odd numbers), such that the superposition is written as
\be
\varphi(x,t) = \frac{A}{\sqrt{2}}\,\left[\sin(K_1\,x)\,\cos(\Omega_1\,t) + \sin(K_3\,x)\,\cos(\Omega_3\,t)\right] \,,
\ee
where $A$ is an amplitude of transverse oscillations.

In $x=L/2$, the wave slope $\vec{\nabla}\varphi(x=L/2,t) = 0$ for any $t$. According to the $\varphi$-guidance formula (\ref{eth_symb}), the concretion can be thus located at this point and remain here -- as for any eigenstate with an odd number $n$.

Let us now  study the energy of the concretion remaining at $x=L/2$. According to Eq.~(\ref{esa_Econc}), after averaging over a transverse oscillation period and when $\omega_n \ll \Omega_m$, the energy of the concretion is 
\be 
\Wconc = \frac{1}{2}\,m_0\,A^2\,\Omega_m^2\left[\left(1+\cos( \Delta\omega\, t)\right)\left(1 + \frac{\omega_1+\omega_3}{2\,\Omega_m}\right)\right] \,,
\ee
where $\Delta\omega = \omega_3-\omega_1$. Rather than calculating the average value of $\Wconc$ over time, it appears that the total energy of the concretion, $\Wconc$, is periodically equal to zero. The analogy with quantum mechanics seems no longer to hold. Furthermore a total energy of the concretion equal to zero does not seem to be realistic, in particular if the `particle' concretion is a simplified representation of a soliton. Then, the toy model suggested in this paper should not be suitable for dealing with the superposition of eigenstates. If more complex considerations demand that the kind of concretion has a total energy near to $m_0\,c_m^2$, we could expect that same phenomena as for walkers appear: the system could exhibit very short transitions between eigenstates~\cite{yc_chaos14}.

%%%%%%%%%%%%%%%%%%%%%%%%%%
%%%%%%%%%%%%%%%%%%%%%%%%%%
%%%%%%%%%%%%%%%%%%%%%%%%%%


\begin{thebibliography}{}
\itemsep=0pt
\parsep=0pt
\leftmargin=\parindent
\itemindent=-\parindent

\bibitem{yc_walking05}Y. Couder, S. Protière, E. Fort, A. Boudaoud, {\it Walking and orbiting droplets.} Nature {\bf 437}, 208 (2005).

\bibitem{bush_review15}J. W. M. Bush, {\it Pilot-Wave Hydrodynamics.} Annu. Rev. Fluid Mech. {\bf 47}, 269-292 (2015).

\bibitem{ldb_ondeguidee1927}L. de Broglie, {\it La mécanique ondulatoire et la structure atomique de la matière et du rayonnement.} J. de Phys. Radium, série \cRM{6}, t. \cRM{8}, n\textsuperscript{o}5 (1927).

\bibitem{yc_wavepartduality11}Y. Couder and E. Fort, {\it Probabilities and trajectories in a classical wave-particle duality probabilities and trajectories in a classical wave-particle duality.} J. of Phys., Conf. Series {\bf 361}, 012001, (2012).

\bibitem{bush_phystoday16}J. W. M. Bush, {\it The new wave of pilot-wave theory.} Physics Today {\bf 68}(8), 47 (2015).

\bibitem{masselotte}C. Borghesi, {\it Dualité onde-corpuscule formée par une masselotte oscillante dans un milieu élastique: étude théorique et similitudes quantiques.} Ann. Fond. de Broglie {\bf 42}(1), 161 (2017). English translation: {\it Wave-particle duality coming from a bead oscillator in an elastic medium, theoretical study and quantum similarities}; arXiv:1609.09260v3 [physics.class-ph]. 

\bibitem{yc_sao1999}A. Boudaoud, Y. Couder, and M. Ben Amar, {\it A self-adaptative oscillator.} Eur. Phys. J. B {\bf 9}, 159-165 (1999).

\bibitem{holland1}P. R. Holland, {\it Hamiltonian theory of wave and particle in quantum mechanics I: Liouville's theorem and the interpretation of the de Broglie-Bohm theory.} Nuovo Cimento B {\bf 116}, 1043-1070 (2001).

\bibitem{holland2}P. R. Holland, {\it Hamiltonian theory of wave and particle in quantum mechanics II: Hamilton-Jacobi theory and particle back-reaction.} Nuovo Cimento B {\bf 116}, 1143-1172 (2001).

\bibitem{holland3}P. R. Holland, {\it Quantum back-reaction and the particle law of motion.} J. Phys. A, {\bf 39}, 559 (2006).%, No. 3

\bibitem{durt_1}T. Durt, {\it Generalized guidance equation for peaked quantum solitons and effective gravity.} Europhys. Lett. {\bf 114}, No. 1 (2016).

%\bibitem{durt_2}T. Durt, {\it Generalized guidance equation for peaked quantum solitons: the single particle case.} Ann. fond. de Broglie (to appear); arXiv:1602.03133v2 [quant-ph].
\bibitem{durt_2}T. Durt, {\it L. de Broglie’s double solution and self-gravitation.} Ann. Fond. de Broglie {\bf 42}(1), 73 (2017).

\bibitem{ldb_tentative}L. de Broglie, {\it Une tentative d'interprétation causale et non linéaire de la mécanique ondulatoire.} Gauthier-Villars Ed., Paris, 1956. English translation: {\it Non-linear Wave mechanics -- A causal interpretation.} Elsevier Ed., Amsterdam, 1960.

%\bibitem{ldb_interpretation1959}L. de Broglie, {\it L'interprétation de la mécanique ondulatoire}. J. de Phys. Radium, t. 20, n\textsuperscript{o}12, 963-979 (1959).
\bibitem{ldb_interpretation1987}L. de Broglie, {\it Interpretation of quantum mechanics by the double solution theory.} Ann. fond. de Broglie {\bf 12}, n\textsuperscript{o}4 (1987). English translation from a paper originally published in the book {\it Foundations of Quantum Mechanics -- Rendiconti della Scuola
Internazionale di Fisica ``Enrico Fermi''}; Course 49 (1970) ed. by B. d'Espagnat, Academic Press N.Y. 1972. 

\bibitem{yc_path-memory-pnas10}E. Fort, A. Eddi, A. Boudaoud , J. Moukhtar and Yves Couder, {\it Path-memory induced quantization of classical orbits}, Proc. Natl. Acad. Sci. USA {\bf 107}, vol. 41, 17515-17520, (2010).

\bibitem{yc_SO-eingenstates14}S. Perrard, M. Labousse, M. Miskin, E. Fort and Y. Couder, {\it Self-organization into quantized eigenstates of a classical wave-driven particle.} Nat. Comm. {\bf 5}, 3219 (2014).

\bibitem{ldb_5dim1927}L. de Broglie, {\it L'univers à cinq dimensions et la mécanique ondulatoire.} J. de Phys. Radium \cRM{6}, t. \cRM{8}, n\textsuperscript{o} 2, (1927).%\no 2

\bibitem{molacek1}J. Mol\'{a}\v{c}ek and J. W. M. Bush, {\it Drops bouncing on a vibrating bath.} J. Fluid Mech. {\bf 727}, 582-611 (2013).

\bibitem{molacek2}J. Mol\'{a}\v{c}ek and J. W. M. Bush, {\it Drops walking on a vibrating bath: towards a hydrodynamic pilot-wave theory.} J. Fluid Mech. {\bf 727}, 612-647 (2013).

\bibitem{fargue}D. Fargue, {\it Louis de Broglie’s ``double solution" a promising but unfinished theory.} Ann. Fond. de Broglie {\bf 42}(1), 9 (2017).

\bibitem{double_sol_90ans}S. Colin, T. Durt and R. Willox, {\it L. de Broglie's double solution program: 90 years later.} Ann. Fond. de Broglie {\bf 42}(1), 19 (2017).

\bibitem{yc_orbiting_source13}E. Fort and Y. Couder, {\it Trajectory eigenmodes of an orbiting wave source.} Europhys. Lett. {\bf 102}, 16005 (2013).

\bibitem{yc_auto_orb16}M. Labousse, S. Perrard, Y. Couder, and E. Fort, {\it Self-attraction into spinning eigenstates of a mobile wave source by its emission back-reaction.} Phys. Rev. E {\bf 94}, 042224 (2016).

\bibitem{Q_th_motion}P. R. Holland, {\it The quantum theory of motion.} Cambridge U. Press, 1995.

\bibitem{applied_bohmian}X. Oriols, J. Mompart, {\it Overview of Bohmian Mechanics.} In X. Oriols, J. Mompart, {\it Applied Bohmian Mechanics: From Nanoscale Systems to Cosmology}; Pan Stanford Ed., 2012.

\bibitem{bohm1}D. Bohm, {\it A Suggested Interpretation of the Quantum Theory in Terms of ``Hidden'' Variables. I.} Phys. Rev. {\bf 85}, 166-179 (1952).

\bibitem{wave-mediated}C. Borghesi, J. Moukhtar, M. Labousse, A. Eddi, E. Fort, and Y. Couder, {\it Interaction of two walkers: Wave-mediated energy and force.} Phys. rev. E {\bf 90}, 063017, (2014).

\bibitem{landau_MQ}L. Landau et E. Lifchitz, {\it Mécanique quantique.} Ed. Mir, Moscow, 1975.

\bibitem{poincare_dyn_e-}H. Poincaré, {\it Sur la dynamique de l'électron.} Rendiconti del Circolo Matematico di Palermo (1905 July 23), 1906. English translation and modernised presentation: H. M. Schwartz, {\it Poincaré's Rendiconti Paper on Relativity}; Am. J. Phys., {\bf 39} 1287-1294 (1971), {\bf 40} 862-871 \& 1282-1287(1972).
%https://en.wikisource.org/wiki/Translation:On_the_Dynamics_of_the_Electron_(July)
%Republished {\it e.g.} in: H. Poincaré, {\it La mécanique nouvelle}; Ed. Jacques Gabay, Paris, 2007. 

\bibitem{pierseaux_RR}Y. Pierseaux, {\it La ``structure fine'' de la relativité restreinte}, L'Harmattan, Paris, 1999. 

\bibitem{zee}A. Zee, {\it Quantum field theory in a nutshell}, Princeton University Press, 2003.

\bibitem{yc_chaos14}S. Perrard, M. Labousse, E. Fort, and Y. Couder, {\it Chaos Driven by Interfering Memory}, Phys. Rev. Lett. {\bf 113}, 104101, (2014).


\end{thebibliography}
\end{document}